\documentclass[
twocolumn,
superscriptaddress,
 amsmath,amssymb,
longbibliography,
]{revtex4-2}
\usepackage{silence}
\usepackage[utf8]{inputenc}
\usepackage[english]{babel}
\usepackage{amsmath}
\usepackage{amssymb}
\usepackage{svg}
\usepackage{graphicx}
\usepackage{tikz}\usetikzlibrary{patterns}\usetikzlibrary{calc,3d}
\usepackage{stix}
\usepackage{placeins}
\usepackage{siunitx}
\usepackage{upgreek}
\usepackage{longtable}
\usepackage{siunitx}
\usepackage{microtype}
\usepackage{svg}\svgsetup{inkscapelatex=false,inkscapearea=page}
\usepackage{adjustbox} 
\usepackage[unicode=true,pdfusetitle,bookmarks=true,bookmarksnumbered=false,bookmarksopen=false,breaklinks=true,pdfborder={0 0 0},pdfborderstyle={},backref=false,colorlinks=false]{hyperref}
\usepackage{standalone}
\usepackage{xspace}
\usepackage{color}

\usepackage{natbib}
\usepackage{bibunits} 

\usepackage{array}
\newcolumntype{H}{>{\setbox0=\hbox\bgroup}c<{\egroup}@{}}

\usepackage[most]{tcolorbox}
\usepackage{ulem,xpatch}

\graphicspath{{figures/main_figures/},{figures/supp_figures/}}
\newcommand{\siliconnitride}{\mathrm{Si_3 N_4}\xspace}
\renewcommand{\unit}[2]{$#1 \, \mathrm{#2}$}
\newcommand{\mathunit}[2]{#1 \, \mathrm{#2}}
\newcommand{\resw}{\Omega_\mathrm{m}\xspace}
\newcommand{\gammam}{\Gamma_\mathrm{m}\xspace}
\newcommand{\gammaba}{\gamma_{\mathrm{qba}}\xspace}
\newcommand{\gammafb}{g_{\mathrm{fb}}\xspace}
\newcommand{\gammasq}{g_\mathrm{s}\xspace}
\newcommand{\xone}{X_1\xspace}
\newcommand{\xtwo}{X_2\xspace}
\newcommand{\Vdc}{V_{\mathrm{DC}}\xspace}
\newcommand{\Vpar}{V_{\mathrm{p}}\xspace}
\newcommand{\kappap}{k_\mathrm{p}\xspace}
\newcommand{\Vth}{V_{\mathrm{th}}\xspace}
\newcommand{\Cder}{C^{''}(x)|_{x_\mathrm{eq}}}
\newcommand{\Omegap}{\Omega_\mathrm{p}}

\newcommand{\tpar}{\left( t \right)}
\newcommand{\ox}{\hat{x}}

\newcommand{\oyt}{\tilde{y}}
\newcommand{\oy}{\hat{y}}
\newcommand{\xzp}{x_\mathrm{zpf}}
\newcommand{\om}{\left(\omega\right)}

\newcommand{\Gopt}{\Gamma_{\mathrm{qba}}}

\newcommand{\Ffbt}{\tilde{{F}}_\mathrm{fb}}

\newcommand{\Gfb}{\Gamma_{\mathrm{fb}}}
\newcommand{\gms}{\gamma_{\mathrm{meas}}}

\begin{document} 

\title{Strong Thermomechanical Noise Squeezing Stabilized by Feedback}

\author{Aida Mashaal}
\thanks{These authors contributed equally to this work.}
\affiliation{Niels Bohr Institute, University of Copenhagen, Blegdamsvej 17, 2100, Copenhagen, Denmark}
\affiliation{Center for Hybrid Quantum Networks, Niels Bohr Institute, University of Copenhagen, Blegdamsvej 17, 2100, Copenhagen, Denmark}
\author{Lucio Stefan}
\thanks{These authors contributed equally to this work.}
\affiliation{Niels Bohr
Institute, University of Copenhagen, Blegdamsvej 17, 2100, Copenhagen, Denmark}
\affiliation{Center for Hybrid Quantum Networks, Niels Bohr Institute, University of Copenhagen, Blegdamsvej 17, 2100, Copenhagen, Denmark}
\author{Andrea Ranfagni}
\affiliation{Niels Bohr Institute, University of Copenhagen, Blegdamsvej 17, 2100, Copenhagen, Denmark}
\affiliation{Center for Hybrid Quantum Networks, Niels Bohr Institute, University of Copenhagen, Blegdamsvej 17, 2100, Copenhagen, Denmark}
\author{Letizia Catalini}
\affiliation{Laboratory for Solid State Physics, ETH Z\"urich, 8093 Z\"urich, Switzerland}
\affiliation{Quantum Center, ETH Z\"urich, 8093 Z\"urich, Switzerland }
\author{Ilia Chernobrovkin}
\affiliation{Niels Bohr Institute, University of Copenhagen, Blegdamsvej 17, 2100, Copenhagen, Denmark}
\affiliation{Center for Hybrid Quantum Networks, Niels Bohr Institute, University of Copenhagen, Blegdamsvej 17, 2100, Copenhagen, Denmark}
\author{Thibault Capelle}
\affiliation{Niels Bohr Institute, University of Copenhagen, Blegdamsvej 17, 2100, Copenhagen, Denmark}
\affiliation{Center for Hybrid Quantum Networks, Niels Bohr Institute, University of Copenhagen, Blegdamsvej 17, 2100, Copenhagen, Denmark}
\author{Eric C. Langman}
\affiliation{Niels Bohr Institute, University of Copenhagen, Blegdamsvej 17, 2100, Copenhagen, Denmark}
\affiliation{Center for Hybrid Quantum Networks, Niels Bohr Institute, University of Copenhagen, Blegdamsvej 17, 2100, Copenhagen, Denmark}
\author{Albert Schliesser}
\email{albert.schliesser@nbi.ku.dk}
\affiliation{Niels Bohr Institute, University of Copenhagen, Blegdamsvej 17, 2100, Copenhagen, Denmark}
\affiliation{Center for Hybrid Quantum Networks, Niels Bohr Institute, University of Copenhagen, Blegdamsvej 17, 2100, Copenhagen, Denmark}

\begin{abstract}
Squeezing the quadrature noise of a harmonic oscillator enhances its sensitivity. Stabilization of the anti-squeezed quadrature can overcome the 3 dB limit of squeezing by parametric modulation. Here, we apply this method to soft-clamped membrane resonators, which hold promise for both classical and quantum sensing applications. We compare piezo and capacitive parametric modulation, and observe thermomechanical squeezing by up to 17 dB and 21 dB, respectively. Finally, we provide a full quantum theory of a combination of this approach with quantum-limited motion measurement and conclude that quantum squeezing is feasible at moderate cryogenic temperatures and realistic device parameters.
\end{abstract}

\maketitle

\section{Introduction}
Mechanical resonators can sense a variety of physical entities---including force \cite{Eichler2022}, acceleration \cite{Gerberding2015}, mass \cite{Chaste2012}, voltage \cite{Bagci2014} and magnetic fields \cite{Simonsen2019}---with astounding sensitivity.
As technical improvements continue to advance their performance, sensors now approach limits dictated by quantum mechanics.
For example, most common techniques to read out a mechanical resonator position add noise that is subject to the standard quantum limit (SQL) \cite{Thorne1992,Aspelmeyer2014}.
However, techniques to overcome the latter have been demonstrated recently \cite{Mason2019, Yu2020}.

In addition, the sensitivity can be limited by fluctuations intrinsic to the mechanical resonator, primarily its thermal fluctuations and then ultimately its vacuum fluctuations (zero-point motion).
%
%
When such limits are reached, a route to improve the sensitivity consists of squeezing the measured mechanical quadrature noise. 
This is possible through parametric modulation of the elastic constant, as demonstrated early on with cantilever force sensors \cite{Rugar1991}.
Recently, this principle has been extended to the quantum regime for single trapped ions \cite{Burd2019}.
 The latter constitutes an impressive conceptual achievement, even if concrete sensing applications are less obvious. Quantum squeezing has also been achieved with superconducting micromechanical resonators cooled to dilution refrigerator temperatures, using microwave reservoir engineering \cite{wollman2015, pirk2015, lecocq2015}.

In contrast, our goal is to demonstrate strong parametric squeezing---ultimately squeezing also quantum noise---in a versatile sensor platform, namely soft-clamped nanomechanical membrane resonators. 
%
%
%
They offer large sensing areas on the order of $(0{.}3~\mathrm{mm})^2$, combined with nanogram effective masses and quality factors that can exceed $10^9$~\cite{Seis2022}.
Since their recent advent \cite{Tsaturyan_2017}, they have already been used for nanoscale force imaging \cite{Halg2021}, measurement of quantum fluctuations of optical forces \cite{Rossi_2018, Huang2024}, and are promising candidates for sensing minute forces e.g. from individual spins \cite{Eichler2022,Kolkowitz2012,Kosata2020}.

Given the usually large thermal motion undergone by such resonators even at cryogenic temperatures, a potential limitation of conventional squeezing protocols is that in the steady state, parametric systems reach an oscillation instability when the target quadrature is squeezed by $\mathunit{3}{dB}$. 
However, this limit can be overcome using e.g. a detuned parametric drive~\cite{Szorkovszky2013}, or by actively stabilizing the anti-squeezed quadrature~\cite{Vinante2013, Pontin2014}. In a proof-of-principle experiment, the latter protocol has been used to demonstrate squeezing of up to $\mathunit{15}{dB}$~\cite{Poot2015}.


In this work, we benchmark a similar protocol applied to our membrane force sensor, and compare the performance of two different parametric modulation methods: one based on piezo actuation and the other on capacitive coupling.
By applying quadrature feedback stabilization, we demonstrate a maximum squeezing of $\mathunit{17}{dB}$ and of $\mathunit{21}{dB}$, respectively, with moderate device parameters at room temperature.
Furthermore, we develop a model that predicts the behavior of the protocol in the quantum regime under continuous quantum-limited monitoring.
We estimate that, with realistic parameters and at a moderate bath temperature $T=\mathunit{10}{K}$, squeezing below the zero-point fluctuations is attainable.

\begin{figure*}
\centering
\includegraphics[width=\textwidth, height=8cm]{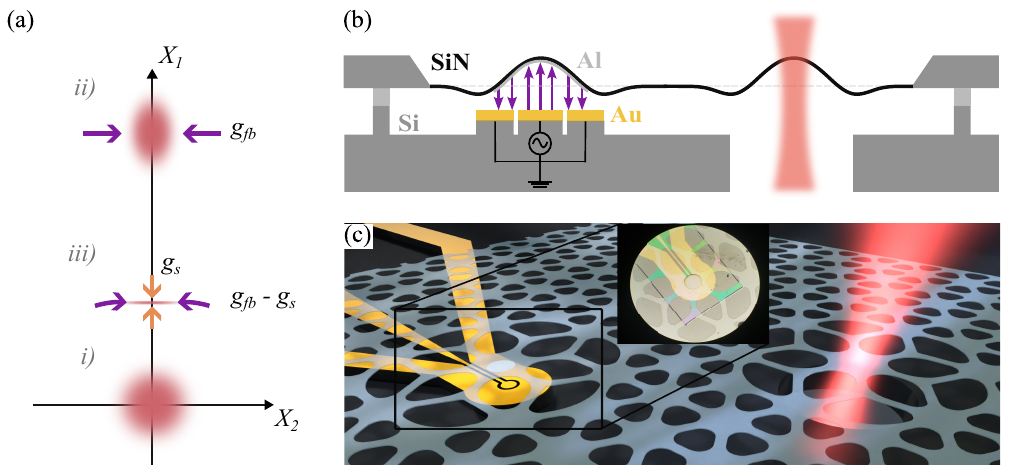}
\caption{Experimental protocol and platform for capacitive parametric driving. (a) A sketch of the experimental protocol implemented to create classical squeezed states of the mechanical motion. A mechanical resonator in a thermal state (i) is displaced along the amplitude quadrature $X_{1}$ by a resonant drive while the phase quadrature $X_{2}$ is stabilized through a phase-locked loop (PLL), as depicted in (ii). The $X_{2}$ quadrature is therefore dampened by a feedback rate  $\Gamma_\mathrm{fb} \propto \gammafb$  (purple arrows). By applying a parametric drive, the mean value of the $X_{1}$ quadrature is deamplified, while the  variance is squeezed by a rate $\Gamma_\mathrm{s} \propto \gammasq$ (orange arrows), shown in (iii). The anti-squeezing of the $X_{2}$ quadrature variance is counteracted by the PLL, preventing the system from undergoing self-oscillation. (b) A double-defect silicon nitride (SiN) membrane lies on aluminum (Al) pillars of a silicon (Si) chip with a printed gold (Au) electrode circuit. An aluminum pad is deposited onto one of the defects, forming a capacitor with the underlying electrodes. For detection purposes, a hole is etched on the side of the electrode chip beneath the non-metallized defect. (c) A rendered angled view of the capacitive platform with an inset of a microscopic image.}
\label{fig:setup}
\end{figure*}

\section{Protocol}

A mechanical oscillator, characterized by an effective mass $m$, 
a mechanical damping rate $\gammam$ and a
natural elastic constant $k_\mathrm{m}$ is driven around resonance by a force $F_\mathrm{d}=F_0 \cos \left( \Omega t  \right)$. The angular frequency $\Omega$ of the drive is set by an electronic oscillator which is phase-locked to the mechanical motion. The parametric drive, offset by a phase $\phi_\mathrm{p}$ relative to the electronic oscillator, modulates the mechanical stiffness at twice the drive frequency $2\Omega$ with an amplitude $k_\mathrm{p}$.

We express the apparent mechanical motion, referenced to the phase-locked electronic oscillator, as \begin{equation}
    \ddot{x} +
    \gammam \dot{x} +
    \frac{1}{m} \left[ k_\mathrm{m} + 
                        \kappap \sin{\left(2 \Omega t + \phi_\mathrm{p} \right)}
                \right] x = 
    \frac{F_\mathrm{th}+F_\mathrm{d}+F_\mathrm{fb}}{m} \; ,
    \label{eq:eqmotion}
\end{equation}
where  $F_\mathrm{th}$ is the thermomechanical noise force and  $F_\mathrm{fb}$ is an effective force representing the action of the phase-locked loop that stabilizes the relative phase between the mechanical oscillation and the electronic oscillator drive. 

For a high quality factor oscillator ($Q = \resw / \gammam \gg 1$) and $\Omega \sim \resw$, the motion is described in the frame rotating at the frequency $\Omega$ in terms of the two slow-varying quadratures $\xone(t)$ and $\xtwo(t)$, defined as:
\begin{equation}
x(t) = \xone(t) \sin\left(\Omega t \right)+ \xtwo(t) \cos\left(\Omega t\right) \; . 
\label{eq_rotatingframe}
\end{equation}

When the parametric modulation and the stabilization are turned off, the oscillator is driven exclusively by the thermomechanical noise and the  drive $F_\mathrm{d}$. The phase quadrature $\xtwo$ as a function of the frequency difference $ \left(\Omega - \resw\right)$ shows a dispersive shape, thus providing an error signal suitable for the frequency stabilization.

With both the feedback and the parametric modulation turned on, the phase space vector---whose components are represented by a sum of a coherent amplitude $\bar{X_i}$ and a fluctuating component $\delta X_i$---
fluctuates around the equilibrium position $\left(\bar{X}_1, \bar{X}_2 \right)^T=\left(F_0/m \,  \resw \gammam \left( 1 + \gammasq \right),0  \right)^T$, where $\gammasq=\Gamma_s / \gammam$ and 
$\Gamma_\mathrm{s} = \kappap / (2 m \resw)$ is the squeezing rate. The equations for the residual fluctuations of the two quadratures in the Fourier domain read~\cite{briant2003}:
\begin{equation}
\begin{aligned}
    & \delta \tilde{\xone}\left(\omega \right) =
    \frac{1}{2 m \resw} \, \frac{1}{-i \omega
                                 + \frac{\gammam}{2} \left(1+\gammasq \right)}
    \tilde{F}_{\mathrm{th}2}\left(\omega \right) \; ,\\
    & \delta \tilde{\xtwo}\left(\omega \right) = 
    - \frac{1}{2 m \resw} \, \frac{1}{-i \omega
                                 + \frac{\gammam}{2} \left(1-\gammasq +\gammafb\right)}
    \tilde{F}_{\mathrm{th}1}\left(\omega \right) \; ,
    \label{eq:freq_quads}
\end{aligned}
\end{equation}    
where $\gammafb = \Gfb / \gammam $, with the feedback force expressed as $F_\mathrm{fb}=m \Omega_\mathrm{m} \Gfb \xtwo \sin{\left(\Omega t\right)}$, and the thermal force $F_{\mathrm{th}} \tpar =F_{\mathrm{th1}} \tpar \sin\left(\Omega t \right)+ F_{\mathrm{th2}} \tpar \cos\left(\Omega t \right)$.
 From Eq.~(\ref{eq:freq_quads}) it follows that the two quadratures have different mechanical susceptibilities. The amplitude quadrature $\xone$, which is in-phase with the parametric modulation, experiences an increased damping rate. Contrarily, as the squeezing increases, the damping of the phase quadrature $\xtwo$ is reduced. The system reaches the instability threshold when $\Gamma_\mathrm{s} = \gammam$, thereby limiting the maximum squeezing of $\delta \bar{X}_1$ to \unit{3}{dB}, unless the feedback is turned on.
 \begin{figure*}
\center
\includegraphics[width=\textwidth]{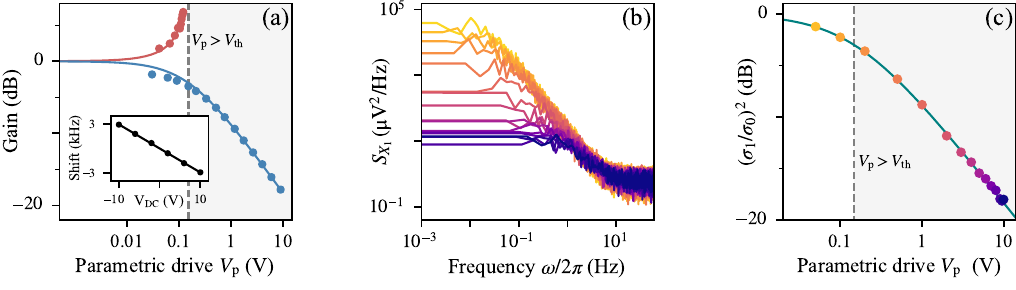}
\caption{Parametric modulation of a membrane resonator using a piezo actuator. (a) Amplitude gain of the $\bar{X}_1$ quadrature as a function of the applied parametric drive, for $\phi_p = \pi/2$ (red dots) and $\phi_p = 0$ (blue dots). When the quadrature is amplified it diverges at the threshold (dashed gray line), while the \unit{3}{dB} limit can be overcome (gray shaded area) when $\xone$ is deamplified. The inset shows the tuning of the resonance frequency as a function of DC piezo voltage. All data are down-sampled for  visual clarity. The solid line are fits to the experimental data. (b) Power spectral density of the $\xone$ quadrature for increasing parametric drive voltage. When the parametric drive voltage is increased from \unit{0}{V} (yellow line) to large values (increasingly dark lines), the total mode damping increases. (c) Variance $\sigma_1^{\; 2}$ obtained from fitting the PSDs in panel (b) normalized by the variance at $\Vpar = \mathunit{0}{V}$, $\sigma_0^{\;2}$. The markers have the same color coding as in panel (b). The solid line is a fit to the experimental data.}
\label{fig:piezo_results}
\end{figure*}

The thermal force quadratures are uncorrelated and the double-sided power spectral density (PSD) reads $S_{F_\mathrm{th1}}=S_{F_\mathrm{th2}} = 2 S_{F_\mathrm{th}}$, with $S_{F_\mathrm{th}}=2 m \gammam k_\mathrm{B} T$.
For the system at equilibrium with the thermal
bath, the variances for the two quadratures are:

\begin{equation}
    \left( \frac{\sigma_1}{\sigma_0} \right)^2 = \frac{1}{1+\gammasq}\; , \qquad 
    \left( \frac{\sigma_2}{\sigma_0} \right)^2 = \frac{1}{1-\gammasq+\gammafb}\; ,
\label{eq:variance_equation}
\end{equation}   
where $\sigma_0^2 = k_\mathrm{B} T / m \, \resw^2$.

As such, the protocol allows to get a displaced squeezed state from an initial thermal state. Thanks to the tracking of the mechanical resonance, it is possible to squeeze the amplitude fluctuations below the standard
3~dB limit.

\section{Experimental Results}
In this work, we employ highly-stressed $\siliconnitride$ soft-clamped membrane resonators patterned with a phononic crystal. The resonators are operated at room temperature and, in order to minimize the effect of gas damping, at a pressure of \unit{10^{-7}}{mbar}. We devise two alternative configurations  to drive the membranes parametrically. In the first configuration, we use a membrane hosting a single defect in the center. The membrane is glued on the four sides of the silicon frame to a ring piezoelectric actuator (Piezomechanik, HPCh 150/15-8/3). When a voltage is applied to the actuator, its vertical expansion (contraction) is accompanied by a  contraction (expansion) in the horizontal plane. As a result, the membrane experiences a modulation of the stress, and therefore of the resonance frequency, via the shear force applied to the frame, resulting in parametric driving. 

The second assembly we explore is illustrated in Fig.~\ref{fig:setup}(b). The membrane hosts a phononic dimer, which gives rise to a symmetric and an anti-symmetric mode whose envelopes span both defects~\cite{Catalini2020}. One of the two defects is metallized with a \unit{50}{nm}-thick aluminum pad, with a diameter of \unit{100}{\upmu m}~\cite{Seis2022}. The membrane is then bonded to a silicon chip with a printed circuit consisting of a circular signal electrode surrounded by a grounding rim. The metallized defect forms a capacitor with the underlying electrodes. A voltage $V(t)$ applied to the circuit creates a modulation of the elastic constant:
\begin{equation}
\kappap =  - \frac{1}{2} \left. \left( \frac{\partial^2 C(x)}{\partial x^2} \right) \right|_{x_\mathrm{eq}} V^2(t) \; ,
\label{eq:kappap_cap}
\end{equation} where $C(x)$ is the total capacitance and $x_\mathrm{eq}$ is the equilibrium position of the membrane.  

The mechanical motion is detected using a Mach-Zender interferometer operating at a wavelength of \unit{1535}{nm}~\cite{Catalini2020}. The interference signal, measured by a balanced photodetector, is demodulated by a lock-in amplifier (HF2LI, Zurich Instruments) at the mechanical resonance frequency. The lock-in amplifier also provides the signals to drive the mechanical resonance and to apply the parametric modulation. 

In order to follow the protocol described in the previous section, the membrane is first driven at resonance. The demodulated signal has a phase offset from the resonant force given by the overall delay caused by the electronics. The resonant force phase is then optimized so that the demodulated quadratures correspond to the phase ($\xtwo$) and amplitude ($\xone$) quadratures. A PLL within the lock-in amplifier provides the feedback needed to stabilize the phase quadrature by adjusting the driving frequency. The latter is frequency-doubled and the resulting signal is sent to the actuator to apply the parametric drive.

\paragraph{Parametric drive with a piezo}
We first study the response of the resonator to the drive scheme shown in Fig.~\ref{fig:setup}(a). The resonator design uses a so called Dahlia-class defect \cite{Rossi_2018} centered in a phononic crystal structure approximately $\mathunit{3.3}{mm} \times \mathunit{3.3}{mm}$ in size, etched into a \unit{41}{nm} membrane of stoichiometric LPCVD $\mathrm{Si_3 N_4}$ with a stress of 1.25 GPa. The resonator has a frequency $\resw = 2\pi \times \mathunit{1.33}{MHz}$ and a quality factor $Q = \left(28 \pm 5\right) \times 10^6$. To certify the quality of the assembly, we first measure the tuning of the resonance frequency with the DC voltage applied to the ring piezo actuator. The shift is linear, with a slope that is typically $\simeq \mathunit{-300}{Hz/V}$ (inset of Fig.~\ref{fig:piezo_results}(a)). 

After turning on the resonant drive and the PLL for the stabilization scheme, we systematically increase the strength of the parametric modulation $\Vpar$ (i.e. the peak voltage of the signal applied to the piezo actuator) while recording the $X_1$ quadrature. The phase of the parametric drive is chosen to get either deamplification  ($\phi_{\mathrm{p}} = 0$) or amplification ($\phi_{\mathrm{p}} = \pm \pi/2$) of the amplitude of $\bar{X}_1$. The resulting amplitude gain curves, with the gain defined as $|\bar{X}_{1}/\bar{X}_{1, \Vpar=0}|$, are plotted in Fig.~\ref{fig:piezo_results}(a). When $\bar{X}_1$ is amplified, the feedback (acting only on $\xtwo$) does not provide stabilization and the quadrature can reach the parametric threshold, which occurs when $\gammasq \equiv \Vpar/\Vth =1 $, where $\Vth$ is the threshold voltage. On the contrary, when the parametric phase is set such that $\xone$ is the quadrature undergoing squeezing, the threshold can be overcome and a maximum quadrature amplitude reduction of $\simeq \mathunit{-17}{dB}$ is observed at $\Vpar = \mathunit{10}{V}$. 
Fitting the measured curves results in an estimated $\Vth = \mathunit{(151 \pm 1)}{mV}$.

We verify the performance of the protocol by directly measuring the squeezing of the thermal fluctuations $\delta \xone(t)$ around its steady-state value $\bar \xone$, as quantified by the variance $\sigma_1^{\; 2}$.
We extract $\sigma_1^{\; 2}$ from the PSD $S_{\xone}(\omega)$, at increasingly larger parametric drive voltage (Fig.~\ref{fig:piezo_results}(b)). Each resonance peak is then fitted with a Lorentzian function while discarding the data that lie in the vicinity of 0 Hz. This corresponds to disregarding the coherent amplitude $\bar \xone$ (and additional low-frequency noise) and evaluating the variance of the fluctuations $\delta \xone$ only. The resulting fit parameters are used to calculate the peak area and hence $\sigma_1^{\; 2}$. The variance as a function of $\Vpar$ is shown in Fig.~\ref{fig:piezo_results}(c).  Fitting the model in Eq.~(\ref{eq:variance_equation}) returns $\Vth = \mathunit{(148 \pm 4)}{mV}$.

Despite the strong squeezing which can be achieved through the piezo actuation, reaching even larger values is experimentally challenging. Applying larger $\Vpar \gg \mathunit{10}{V}$ is not a viable option, due to increased noise levels of the detected mechanical motion, likely attributable to heating of the actuator. Nonetheless, on few assemblies we observed an increase in the squeezing performance---corresponding to a four-fold drop in  $\Vth$---when performing the measurement at specific piezo offset voltage $\Vdc$. We attribute this enhancement to the presence of an in-plane mode with angular frequency $\Omega_\mathrm{ip}(\Vdc) \approx 2 \resw$ (see Supplementary Information).

\begin{figure}
\center
\includegraphics[width=\columnwidth]{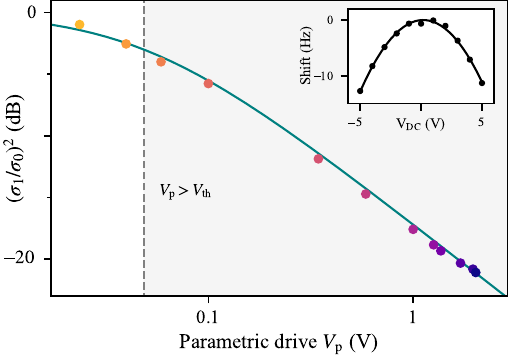}
\caption{Parametric modulation using a capacitive drive. Normalized variance of the amplitude quadrature $(\sigma_1 / \sigma_0)^2$ as a function of the parametric drive voltage, where $\sigma_0 ^{\; 2}$ is the variance at $\Vpar=0$, using the same protocol described in Fig.~\ref{fig:piezo_results} and in the text.  The inset shows the frequency tuning of the mode as a function of the bias voltage $\Vdc$.  The solid lines are fits to the experimental data. }
\label{fig:capacitor_results}
\end{figure}

\paragraph{Capacitive parametric drive} Next, we analyze the performance of the capacitor-based design. For this, we use a $\mathunit{4.8}{mm} \times \mathunit{5.4}{mm}$ Lotus-class dimer membrane~\cite{Planz23}, with a thickness of \unit{50}{nm}. The dimer symmetric mode has a resonance frequency $\resw = 2\pi \times \mathunit{1.3} {MHz}$, with $Q=(0.67\pm0.04) \times 10^6$. We first characterize the frequency shift as a function of an applied DC voltage. The capacitor has a nonlinear spring-softening effect (Eq.~\ref{eq:kappap_cap}), leading to a change of  frequency $ \tilde{\Omega}_0\left(V\right) = \resw \sqrt{1- k_\mathrm{m} \, \Cder \, V^2/2}
$ as shown in the inset of Fig.~\ref{fig:capacitor_results}, where $x_\mathrm{eq}$ is the equilibrium position of the membrane at a given bias voltage $\Vdc$.

In order to generate squeezing, we apply a bias voltage $V(t) = \Vdc + \Vpar \cos(\Omegap t + \varphi)$. The application of non-zero $\Vdc$ is required to reach a low threshold value, since $\Vth = 2 k_\mathrm{m} / (Q \, \Vdc \, \Cder )$~\cite{Bothner2020}. Experimentally, we set the bias voltage to $\Vdc = \mathunit{8}{V}$.
Additionally, we observe that applying a bias voltage increases the quality factor to a value $Q(V) \approx 2 \, Q(\Vdc =0)$, at the chosen bias voltage. 

The variance of $\xone$ at different 
 parametric drive amplitudes (Fig.~\ref{fig:capacitor_results}) is obtained  following the same steps described in the previous paragraph. Despite the  considerably lower $Q$-factor compared to the resonator used to study the piezo-actuated drive, the fitted threshold voltage is almost one order of magnitude less, with a value $\Vth =\mathunit{ (39 \pm 3)}{mV}$. This results in a maximum observed squeezing of the variance $(\sigma_1 / \sigma_0)^2 = \mathunit{-21}{dB}$. 

Here we remark that the observed value is currently limited by experimental factors, such as low signal-to-noise ratio, heating due to leakage currents in the electrodes, and low mechanical quality factor. Simple improvements to the interferometer sensitivity and to the device fabrication and assembly are predicted to yield significantly larger squeezing.

\section{Outlook on quantum squeezing}

\begin{figure}
\center
\includegraphics[width=\columnwidth]{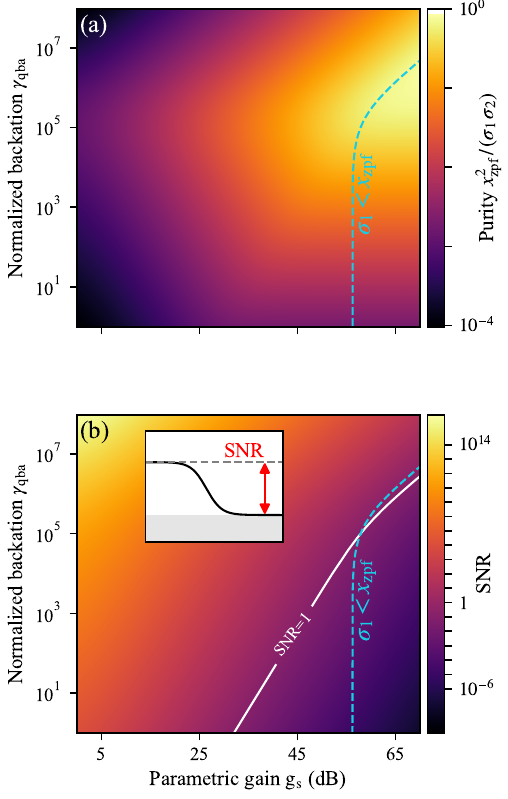} \caption{Prospect for purity and signal-to-noise ratio (SNR) of the detection. The simulations assume $T=\mathunit{10}{K}$, $\resw = 2\pi \times \mathunit{1}{MHz}$, $Q=10^9$, $\eta_\mathrm{det}=0.77$.
(a) Purity of the mechanical state at the optimal feedback which minimizes $\sigma_2^2$, as a function of the  normalized squeezing rate and the backaction rate. In the region to the right of the light-blue-dashed line the expected amplitude standard deviation is below the quantum
zero point fluctuations.
(b) Signal-to-noise ratio of the detected $\xone$ quadrature, defined as the ratio of the amplitude of the mechanical peak   ($S_{\xone} \left(0 \right)$, inset)  and the imprecision noise floor ($S_\mathrm{n n} = \xzp^2 / 2 \Gamma_\mathrm{meas}$), as a function of the normalized squeezing rate and back action rate. The white line marks the contour at which $\mathrm{SNR}=1$.}
\label{fig:theory} 
\end{figure} To assess the experimental feasibility of reaching---and observing---quantum squeezing, we consider  a resonant, high
finesse optical cavity operating in the bad cavity regime, characterized by a linewidth $\kappa/2\pi \gg \Omega_\mathrm{m}/2\pi$. 
Monitoring the output of a resonantly-driven cavity allows 
measuring mechanical displacements characterized by a rate $\Gamma_\mathrm{meas}=\eta_\mathrm{det} 4  g^2/\kappa$, where $g$ is the optomechanical coupling strength \cite{Aspelmeyer2014}, $\kappa$ is the cavity intensity decay rate, $\xzp=\sqrt{\hbar / (2 m \resw})$ and $\eta_{\mathrm{det}}$ is the detection efficiency. The quantum noise of the probe has two detrimental effects on the performance. Its amplitude fluctuations lead to the quantum back action force with a (double sided) power spectral density given by $S_{\mathrm{qba}}\om=\hbar^2 \Gopt/\xzp^2 $ where $\Gopt=4 g^2/\kappa$. The detected signal, referred to the mechanics, is $\oy \tpar =\ox \tpar + \ox_{\mathrm{imp}} \tpar$. The quantum uncertainty in the phase of the probe results in an imprecision noise floor in the detection ($\delta \oy = \ox_{\mathrm{imp}}$), not correlated with the mechanics, which the feedback loop attempts to cancel by actuating the mechanical motion (see Supplementary Information). This results in a force on the phase quadrature in the Fourier domain:
\begin{equation}
    \delta \Ffbt \om = m \Omega_\mathrm{m} \Gfb \delta \oyt_2 \om, 
\end{equation}
where the quadrature of the detected noise signal $\delta \oyt_2$ is defined with the same convention as in Eq.~(\ref{eq_rotatingframe}). 

Adding the backaction of the feedback and of the optical readout in Eq.~(\ref{eq:freq_quads}), the variances of the two quadratures read: \begin{equation} \begin{aligned} \sigma_1 ^2 &=\frac{\xzp^2}{1+\gammasq}\left[\left(2\bar{n}+1\right)+2 \gammaba\right], \\
\sigma_2 ^2 &=\frac{\xzp^2}{1-\gammasq+\gammafb}\left[\left(2 \bar{n}+1\right)+\frac{1}{2}\left(\frac{\gammafb^2}{4 \gms }+4 \gammaba\right)\right],
\label{eq:quantumquads}
\end{aligned}
\end{equation}
where $\gms=\eta_\mathrm{det} \gammaba$, $\gammaba=\Gopt/\gammam$  and $\bar{n}$ is the average occupancy of the thermal bath. 

Through Eq.~(\ref{eq:quantumquads}), we can estimate the expected purity of the state as a function of the quantum back action and the parametric gain (Fig.~\ref{fig:theory}(a)) for realistic device parameters \cite{Rossi_2018}.
For each value of $\gammaba$ and $\gammasq$, the feedback
gain is tuned to minimize $\sigma_2^2$.
The purity approaches $\sqrt{\eta_\mathrm{det}}$ in the
limit of strong squeezing and fast measurement rate.

The expected detection signal-to-noise ratio (SNR) for the amplitude quadrature is reported in Fig.~\ref{fig:theory}(b). In the limit of strong
measurement rate, where the mechanical motion is dominated 
by the quantum back action, it is possible to squeeze the amplitude quadrature below $\xzp$ and detect it with a unity signal-to-noise ratio (SNR) above the imprecision noise.

\section{Conclusions}
By applying active stabilization of the anti-squeezed quadrature, we have shown that both piezo- and capacitively-actuated parametric driving can be used to squeeze the mechanical motion far beyond the $\mathunit{3}{dB}$ limit, reaching $\mathunit{17}{dB}$ and $\mathunit{21}{dB}$, respectively. Improvements of the device design in the case of piezo actuation are not expected to vastly improve the maximum parametric gain. In contrast, we envision that in the case of capacitive drive, realistic improvements---in terms of larger mechanical quality factor, smaller membrane-electrode separation, and reduced leakage current within the circuit---will result in much larger attainable squeezing, with a scaling of $\sigma_1^2\propto d_0^3/V_\mathrm{DC}V_\mathrm{p} Q $ for $\gammasq\gg 1$. Under such modifications, the device will remain compatible with  force sensing applications \cite{Halg2021}.

We also provide a theoretical model to predict the behaviour of the system in the quantum regime and to discuss the feasibility of squeezing below the zero-point motion. At a moderate bath temperature of $\mathunit{10}{K}$, quantum squeezing requires thermomechanical squeezing by more than $\mathunit{56}{dB}$. We find this can be reached with conservative device parameters ($d_0=1\,\mathrm{\mu m}$, $m_\mathrm{eff}=30\,\mathrm{ng}$, $\Omega_\mathrm{m}=2\pi \times\, 1\,\mathrm{MHz}$ and $Q=10^9$) with voltages below the threshold for (e.g., pull-in) instability.

\section{Acknowledgements}
We thank Emil Zeuthen and Jan Ko{\v s}ata for fruitful discussions. This work was supported by the European Research Council project PHOQS (Grant No. 101002179), the Novo Nordisk Foundation (Grant No. NNF20OC0061866), as well as the Independent Research Fund Denmark (Grant No. 1026-00345B). L. S. acknowledges financial support  from the European Union’s Horizon 2020 research and innovation programme under the Marie Skłodowska-Curie grant agreement No 101063285.

\bibliographystyle{apsrev4-2}
\bibliography{references}

\end{document}



\clearpage
\onecolumngrid
\setcounter{equation}{0}
\setcounter{figure}{0}
\setcounter{table}{0}
\setcounter{page}{1}

\phantomsection
\addcontentsline{toc}{title}{Supplementary}
\setcounter{section}{0}

\makeatletter
\renewcommand{\thepage}{S\arabic{page}}
\thispagestyle{plain}
\pagestyle{plain}

\renewcommand{\theequation}{S\arabic{equation}}
\renewcommand{\thefigure}{S\arabic{figure}}
\renewcommand{\bibnumfmt}[1]{[S#1]}
\renewcommand{\citenumfont}[1]{S#1}
\renewenvironment{widetext}{}{}

\newcommand{\om}{\left(\omega\right)}
\newcommand{\omp}{\left(\omega'\right)}
\newcommand{\mom}{\left(-\omega\right)}
\newcommand{\tpar}{\left(t\right)}

\newcommand{\ox}{\hat{x}}
\newcommand{\oxt}{\tilde{x}}

\newcommand{\oy}{\hat{y}}
\newcommand{\oyt}{\tilde{y}}

\newcommand{\oX}{\hat{X}}
\newcommand{\oXt}{\tilde{X}}

\newcommand{\Gm}{\Gamma_{\mathrm{m}}}
\newcommand{\Gmeas}{\Gamma_{\mathrm{meas}}}
\newcommand{\Gs}{\Gamma_{\mathrm{s}}}
\newcommand{\Gfb}{\Gamma_{\mathrm{fb}}}
\newcommand{\Gqba}{\Gamma_{\mathrm{qba}}}
\newcommand{\Gms}{\Gamma_{\mathrm{meas}}}
\newcommand{\gs}{g_{\mathrm{s}}}
\newcommand{\gms}{g_{\mathrm{meas}}}
\newcommand{\gfb}{g_{\mathrm{fb}}}
\newcommand{\gba}{g_{\mathrm{ba}}}
\newcommand{\Gtilde}{\tilde{\Gamma}}
\newcommand{\gammasq}{g_\mathrm{s}\xspace}

\newcommand{\Go}{\Gamma_{1}}
\newcommand{\Gtw}{\Gamma_{2}}

\newcommand{\xzp}{{x}_\mathrm{zpf}}

\newcommand{\Ffb}{\hat{{F}}_\mathrm{fb}}
\newcommand{\Fth}{\hat{{F}}_\mathrm{th}}
\newcommand{\Fqba}{\hat{{F}}_\mathrm{qba}}
\newcommand{\Fqbac}{{F}_\mathrm{opt}}
\newcommand{\Fd}{{F}_\mathrm{d}}
\newcommand{\Fo}{\hat{{F}}}
\newcommand{\Fopt}{\hat{{F}}_\mathrm{opt}}

\newcommand{\Ffbt}{\tilde{{F}}_\mathrm{fb}}
\newcommand{\Ftht}{\tilde{{F}}_\mathrm{th}}
\newcommand{\Fqbat}{\tilde{{F}}_\mathrm{qba}}
\newcommand{\Fdt}{\tilde{{F}}^_\mathrm{d}}
\newcommand{\Fot}{\tilde{{F}}}

\newcommand{\Ft}{\tilde{\mathrm{F}}}

\newcommand{\Yt}{\tilde{\mathrm{Y}}}

\newcommand{\Xo}{\hat{X}_{1}}
\newcommand{\Xt}{\hat{X}_{2}}

\newcommand{\Xot}{\tilde{X}_{1}}
\newcommand{\Xtt}{\tilde{X}_{2}}

\newcommand{\xo}{\hat{x}_{1}}
\newcommand{\xt}{\hat{x}_{2}}

\newcommand{\xzpf}{x_{\mathrm{zpf}}}

\newcommand{\oa}{\hat{a}}
\newcommand{\oac}{\hat{a}_{\mathrm{c}}}
\newcommand{\oad}{\hat{a}^\dagger}
\newcommand{\oadc}{\hat{a}^\dagger_{\mathrm{c}}}
\newcommand{\oat}{\tilde{a}}
\newcommand{\oatd}{\tilde{a}^\dagger}

\newcommand{\Dpar}{\Delta_{\mathrm{par}}}
\newcommand{\Dshift}{\Delta_{\mathrm{shift}}}
\newcommand{\Temp}{T_{\mathrm{bath}}}
\newcommand{\kB}{k_{\mathrm{B}}}
\newcommand{\const}{c}
\newcommand{\meff}{m_{\mathrm{eff}}}

\newcommand{\deteff}{\eta_{\mathrm{det}}}

\begin{center}
\textbf{\large Supplementary Information: Strong Thermomechanical Noise Squeezing Stabilized by Feedback}
\end{center}

\section{Experimental details}

\subsection{Membrane Resonators}

\begin{figure*}[b]
\center
\includegraphics[width=\linewidth]{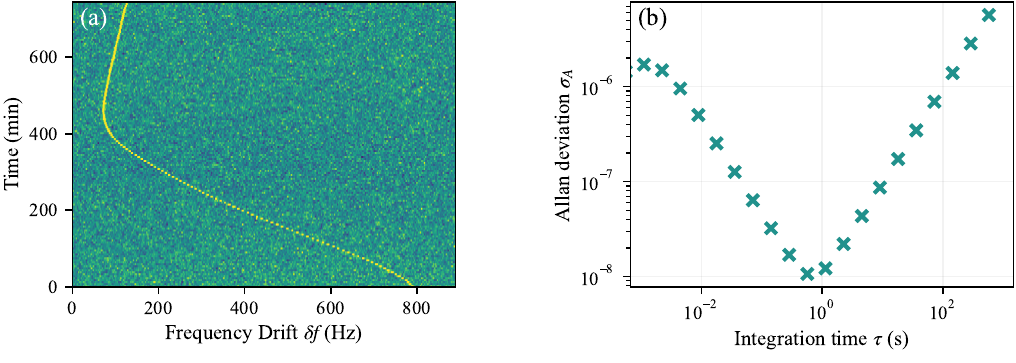}
\caption{Frequency instability measurements. (a) Mechanical frequency drift over an extended period of time. (b) Closed-loop Allan deviation measurement with a minimum deviation at integration times $\tau_\mathrm{min} \sim$ 1 s.} 
\label{fig:freq_ins}
\end{figure*}

Instability of the mechanical frequency $f_{0}$ was observed while optimizing the experimental protocol for achieving thermomechanical squeezing. A tracking of the free-running mechanical frequency can be seen in Fig. \ref{fig:freq_ins} left with a frequency drift of $\sim 100$Hz in the first hour. This drift is significantly larger than the linewidth of the high-Q membrane resonators used in this paper. We here provide an exemplary measurement of frequency instability by means of a closed-loop Allan deviation measurement \cite{Besic2023}. The Allan deviation is defined as
\begin{equation}
\sigma_{A} = \sqrt{\frac{1}{2(N-1)}  \sum_{n=1}^{N-1}\left(\frac{\bar{f}_{n+1} - \bar{f}_{n}}{f_{0}}\right)^2},
\end{equation}
where $f_{n}$ denotes the n-th of $N$ frequencies averaged over an integration time $\tau$, and $f_0$ is the oscillator center frequency. 

We perform the measurement by employing a lock-in amplifier to track the mechanical frequency through a phase-locked loop (PLL). We set the demodulation bandwidth to 3598 Hz and the sampling rate to 28784 Sa/s, as in \cite{Sadeghi2020} for direct comparison purpose, and the PLL bandwidth to 2~Hz. The results of the measurement can be seen in Fig. \ref{fig:freq_ins} (b).

For integration times longer than $\tau_\mathrm{min} \approx 1~\mathrm{s}$, the drift also seen in Fig.~\ref{fig:freq_ins} (a) dominates, leading to a continuously growing error with a slope proportional to the integration time $\tau$. These frequency drifts can limit parametric squeezing measurements, which typically takes from a couple of minutes to hours. We attribute the slow fluctuations to temperature changes within the setup.
For shorter averaging times $\tau < \tau_\mathrm{min}$, the Allan deviation falls with $\tau^{-1}$. This is incompatible with thermomechanical noise, for which we expect an Allan deviation $\propto \tau^{-1/2}$, as seen in \cite{Sadeghi2020}. Instead, this noise is likely due to amplified detection noise \cite{Besic2023}, but its detailed investigation is beyond the scope of this work.

\subsection{Interferometric Detection}

\begin{figure*}
\center
\includegraphics[scale=1.8]{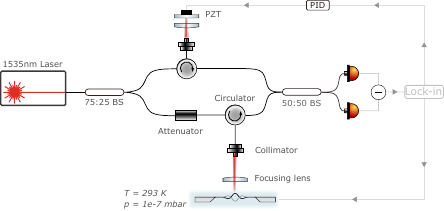}
\caption{Detailed experimental setup. BS: beam splitter. Fiber connections are represented by black lines while electrical connections are drawn in grey. All the fiber components  are polarization-maintaining.}
\label{fig:setup}
\end{figure*}

The experimental setup for measurements reported in this paper is based on earlier work \cite{Catalini2020} and shown in Fig. \ref{fig:setup}. It consists of a Mach-Zehnder interferometer, employed to measure the membrane resonator displacement. A 1550 nm laser is split through a polarization-maintaing beamsplitter into a local oscillator (LO) beam arm and a probe-beam (PB) arm. Through a fiber circulator, the PB travels towards the membrane resonator, held in a vacuum chamber at room temperature. The light enters through port 1 of the fiber circulator and into port 2 to a free-space fiber collimator. The light polarization of the PB is corrected by a polarizer and a half-waveplate to maximize the collection of the light reflected by the membrane. The reflected light is recombined with the LO through a 50:50 beamsplitter onto a balanced photodetector. The motion of the membrane is mapped onto a phase modulation of the PB which is extracted by the homodyne signal fed into a lock-in amplifier. In order to stabilize the homodyne detector to measuring the PB's phase quadrature, the photocurrent is filtered with a proportional-integral-derivative (PID) amplifier and fed back to a piezoelectric transducer displacing a mirror in the optical LO arm.

All the signals within the setup are generated using the lock-in amplifier. This includes the parametric and resonant drive of the membrane, as well as a phase-locked loop (PLL) for stabilizing the system and a PID control to lock the path length of the LO arm. The estimated noise level of this interferometer at typically employed PB power levels is 700 $\mathrm{fm/\sqrt{Hz}}$.

\subsection{Piezo Drive: Observation of Parametric Resonance}

In the case of parametric driving with a piezo, the effect relies on  expansion (contraction) of the ring piezo's radius. As the membrane frame is glued to the piezo, this modulates the membrane's in-plane stress and strain.
We note, however, that the membrane's in-plane motion also supports resonances (in-plane modes).
In this scenario, a elastic continuum model \cite{Catalini2022} predicts that the parametric modulation $k_{p}$ experienced by the out-of-plane mode is set by the spatial overlap between the in-plane and the out-of-plane mode as well as the displacement amplitude of the in-plane mode. 

This implies  that if the parametric drive frequency $\Omega_\mathrm{p}$ is tuned to twice the frequency of the out-of-plane mode $\Omega_{0}$, and simulatenously resonant with an in-plane mode at frequency $\Omega_\mathrm{ip}$, that is
\begin{equation}
2\Omega_{0}=\Omega_\mathrm{p}=\Omega_\mathrm{ip},
\label{S1_eq}
\end{equation}
the parametric modulation is enhanced for the same parametric drive strength.
Such resonant enhancement of the piezo parametric drive due to an in-plane mode may occur coincidentally in a specific sample.
On the other hand, we expect to be able to tune out-of-plane versus in-plane resonance frequencies by applying a static in-plane stress, which only affects out-of-plane resonance frequencies (as confirmed by FEM simulations). 
Thereby, it should be possible to tune through the drive resonance described by the condition of eq.~(\ref{S1_eq}).

We attempt to verify this model by tuning the out-of-plane mode of a Dahlia  generation 2 membrane with a frequency of $\Omega_{0} \approx 2 \pi \cdot 1.33~\mathrm{MHz}$, by applying a DC voltage to the piezo. 
The voltage range accessible with the available instrumentation was $\pm$ 25 V. 
For each applied voltage, we extract a frequency shift of the out-of-plane mode, $\delta \Omega = \Omega - \Omega_{0}$, by  comparing the peak position in a thermal spectrum.

Furthermore, for each applied voltage, we perform an “amplitude sweep measurement”. 
This involves driving the out-of-plane mode simultaneously with a resonant drive at $\Omega_{0}$ and a parametric drive at $\Omega_\mathrm{p}=2\Omega_{0}$ while sweeping the parametric drive strength.
We take two amplitude sweeps at a time and average them. Additionally, we apply feedback cooling to maintain a $Q \approx$ 1M. 
We then determine the parametric drive strength $V_\mathrm{th}$ (in units of voltage), at which the parametric gain reaches the oscillation threshold.

\begin{figure*}
\center
\includegraphics[width=\linewidth]{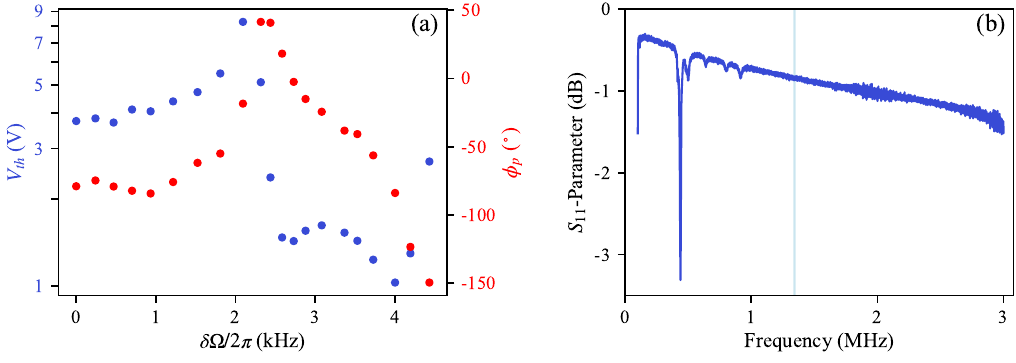}
\caption{Observation of parametric resonances. (a) Threshold voltage $V_{th}$ as a function of frequency detuning $\delta \Omega$. (b) Measured $S_{11}$ reflection spectrum of the piezoelectric actuator. Blue shaded area corresponds to the frequency range of Fig. (a). See text for more details.}
\label{fig:par_res}
\end{figure*}

The results can be seen in Fig. \ref{fig:par_res} (a). 
A variation of the $V_\mathrm{th}$ by a factor of 4 can be observed, as the out-of-plane frequency is tuned. 
This could indicate the crossing of an in-plane mode resonance. 
Moreover, the phase (between dirct and paramatric drive) at which we observe the maximum parametric de-amplification changes. This could be attributed to the phase response of the harmonic in-plane motion. 
As a complementary sanity check, we measure the frequency response of the piezo actuator glued to the membrane. 
The result can be seen in Fig. \ref{fig:par_res} (b).
We do not observe any resonances in the frequency window of interest, ruling out that piezo resonances are the origin of the observed parametric resonance.

The results shown in Figure \ref{fig:par_res} (a) are consistent with the model presented.
Two other membranes have also featured a variation of $V_\mathrm{th}$ when tuning the DC voltage. 
However, within the available frequency tuning range of 6 kHz from the resonance frequency, no parametric resonance (mode crossing) as described by eq.~(\ref{S1_eq}) could be observed.
Further measurements of  $V_\mathrm{th}$ vs detuning in  five additional membranes, revealed no expected correlation. The relatively narrow detuning range of 6 kHz is a significant experimental limitation, given that the simulated average in-plane mode density is as low as 1/(30 kHz). The high Q-factor of the membrane modes further enhances the difficulty of observing the parametric resonance.

In conclusion, whereas we have observed a behavior compatible with a parametric double (drive and pump) resonance due to an in-plane mode in one membrane, our current instrumentation did not allow us to establish this as a reproducible phenomenon.

\section{Electrode simulation and design}
\begin{figure*}
\centering
\includegraphics[width=\textwidth]{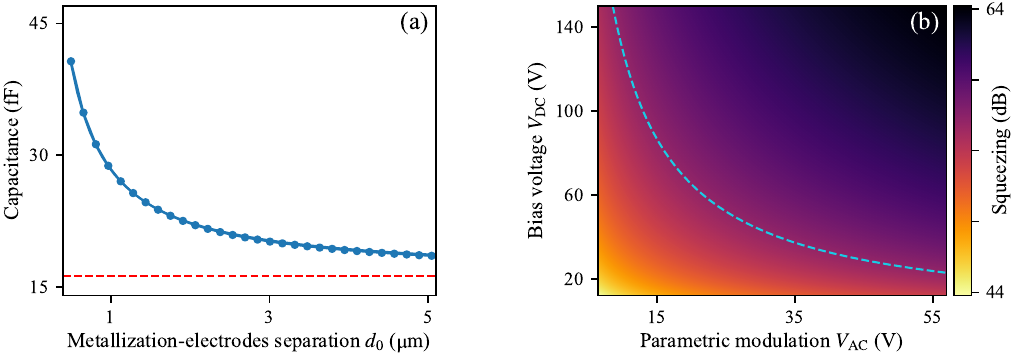}
\caption{Simulations for the capacitive driving. (a) Capacitance of the metallization-electrodes system as a function of the separation $d_0$. At large $d_0$, the value of the capacitance settles to a value of $\mathunit{16}{fF}$ (red dashed line), corresponding to the self-capacitance of the electrodes. (b) Predicted squeezing as a function of bias voltage and parametric modulation amplitude, for $d_0 = \mathunit{1}{\upmu m}$, $m_\mathrm{eff}=\mathunit{30}{ng}$, $\Omega_{\mathrm{m}} = \mathunit{2\pi \times 1}{MHz}$, $Q = 10^9$. The light-blue line marks the contour at which the squeezing reaches $\mathunit{56}{dB}$.}
\label{fig:cap_sims}
\end{figure*}

The electrode geometry is optimized with the aid of COMSOL simulations. To maximize the overall force applied to the capacitor, the design follows a concentric geometry, where the ``live'' electrode surrounded by the ground electrode. The inner conductor diameter is set to $\mathunit{50}{\upmu m}$. The outer diameter of the ground electrode is $\mathunit{270}{\upmu m}$, much larger than the metallization diameter. The gap between the inner conductor and the ground electrode is set to $\mathunit{10}{\upmu m}$,  a trade-off between achieving high overlap between the electrode surface and the metallization, and minimizing the leakage current through the silicon substrate. The separation-dependent capacitance $C(x)$ is extracted numerically from the Maxwell capacitance matrix (Fig.~\ref{fig:cap_sims}(a)). The behavior of $C(x)$ is well approximated by:
\begin{equation}
    C(x) = \frac{\alpha}{x-d_0} + C_0
    \label{eq:cap_dist}
\end{equation}
where $C_0$ is the self-capacitance of the electrodes, $d_0$ is the initial metallization-electrodes separation, and $\alpha$ is a constant encapsulating the geometry of the metallization-electrodes system and the vacuum dielectric permeability. With the current simulation parameters, $\alpha = \mathunit{12}{pF \cdot nm}$, while $C_0 = \mathunit{16}{fF}$.
Using Eq.~(\ref{eq:cap_dist}), we can numerically extract the parametric threshold at $d_0$ and at a given bias $V_{\mathrm{DC}}$, working in the lumped-element description of the resonator. By assuming a conservative set of parameters ($d_0 = \mathunit{1}{\upmu m}$, $m_\mathrm{eff}=\mathunit{30}{ng}$, $\Omega_{\mathrm{m}} = \mathunit{2\pi \times 1}{MHz}$, $Q = 10^9$), we extract the dependence of the squeezing as a function of $V_\mathrm{DC}$ and the parametric modulation amplitude $V_\mathrm{AC}$, as show in Fig.~\ref{fig:cap_sims}(b). With this set of parameters, attaining a squeezing $>\mathunit{56}{dB}$  with realistic values of $V_\mathrm{DC}$ and $V_\mathrm{AC}$ is possible, and can be further improved by reducing $d_0$. However, here we also remark that, to attain more accurate prediction of the parametric instability threshold, a model keeping into account of the spatial dependence of $C(x)$ and the overlap between mode shape and metallization needs to be developed.

\section{Theoretical derivations}

\subsection{Model}
To evaluate the quantum constraints
in the protocol performance,
we consider the scenario 
where a resonant, high
finesse optical cavity operating in the bad cavity regime
is exploited
as quantum detector to probe the
mechanics.
We consider the simplified case 
where the feedback loop is ideal 
(infinite bandwidth of the electronic response) and optimized so that
the electrical drive angular frequency ($\Omega$)
matches that of the oscillator ($\Omega_\mathrm{m}$).

We express the apparent mechanical motion, referenced to the phase-locked electronic oscillator as:
\begin{equation}
\ddot{\ox}+\Gm \dot{\ox} + \frac{1}{m} \left[k_\mathrm{m}+ k_\mathrm{p} \sin \left(2 \Omega t\right)\right]\ox=\frac{\Fth+\Fopt+\Ffb+\Fd }{m}, \label{eq_motion}
\end{equation} 
where $\Gm$ is the intrinsic damping rate and $m$ the effective mass of the oscillator. The stiffness of the 
oscillator ($k_\mathrm{m}$) is modulated at twice the oscillator frequency, with an amplitude $k_\mathrm{p}$. The oscillator is driven by the thermal force $\Fth$, the optical force $\Fopt$
exerted by the cavity to read out the mechanical position and the effective force  $\hat{F}_\mathrm{fb}$, representing the action of the phase-locked loop that stabilizes the relative phase between the mechanical oscillation and the electronic oscillator's drive $\Fd =F_\mathrm{0}   \cos\left(\Omega t\right)$. 

The evolution of the cavity mode field operator 
($\oa$)  in the frame rotating at the laser frequency, in the linearized description of cavity 
optomechanics, reads \cite{Aspelmeyer2014}:
\begin{equation}
\dot{\oa}= - \frac{\kappa}{2} \oa + i \left( \Delta + g_0 \frac{\ox}{\xzp}\right)\oa+ \sqrt{\kappa}\oa_\mathrm{in},
\label{eq_opt_field}
\end{equation} 
where  $\kappa$ is the intensity decay rate, $\Delta$ is the angular frequency difference between the laser and the cavity mode, $\oa_\mathrm{in}$ represents the field that enters the cavity,  sum of a coherent drive $\alpha_\mathrm{in}$ (real valued, without loss of generality)  and
the vacuum fluctuations $\delta \oa_\mathrm{in}$.
$g_0$ is the vacuum optomechanical coupling strength and $\xzp=\sqrt{\frac{\hbar}{2 m \Omega_\mathrm{m}}}$. 

The average radiation pressure force exerted by 
the cavity field $\langle \Fopt \rangle=\frac{\hbar g_0 \left\langle \oad \oa \right \rangle}{\xzp}$ 
leads to a displacement of the oscillator 
equilibrium position  equal to $x_0=\frac{\langle \Fopt \rangle }
{m \Omega_\mathrm{m}^2}$.
The oscillator motion is driven at resonance by the 
 electric force $\Fd$, 
which leads to the coherent oscillation
$x_d (t)=\frac{F_0  \sin\left(\Omega t\right)}{m \Omega_\mathrm{m} \Gm \left(1+ \gs \right)}$, where $\gammasq=\Gamma_s / \Gm$ and 
$\Gamma_\mathrm{s} = k_\mathrm{p} / (2 m \Omega_\mathrm{m})$ is the squeezing rate.
The shift $\langle \ox \rangle =x_0+x_d $ is
omitted in the subsequent discussion, after the 
 redefinition of the cavity detuning $\Delta'=\Delta+\langle \oa \rangle x_0$, which is set to zero in the 
 following, since the cavity is used as a probe.

The equation (\ref{eq_motion}) can be recast for the noisy
component of the position $\delta \ox$ (with $\ox= \langle \ox \rangle + \delta \ox$):
\begin{equation}
\delta\ddot{\ox}+\Gm \delta\dot{\ox} + \frac{1}{m}\left[k_\mathrm{m}+ k_{\mathrm{p}} \sin \left(2 \Omega t\right)\right]\delta\ox=\frac{\Fth+\Fqba+\Ffb}{m}, \label{eq_motion_1}
\end{equation} 
where $\Fqba$ is the quantum radiation pressure force exerted by the cavity field ($\Fopt=\langle \Fopt \rangle + \Fqba$).

For an oscillator with a high quality factor, the 
phase space dynamic is conveniently described in a frame rotating at
the angular frequency $\Omega$ 
\begin{equation}
\delta \ox \tpar= \delta \Xo \tpar \sin\left(\Omega t\right)+  \delta \Xt \tpar \cos\left(\Omega t\right),\label{quad_def}
\end{equation} 
where $\delta \Xo \tpar$ and $ \delta \Xt \tpar$ are the slowly varying amplitude and phase quadratures fluctuations
of the mechanics as defined in the main text, respectively.

The feedback force can be written as $\Ffb=m \Omega_\mathrm{m} \Gfb \oy_2  \sin \left(\Omega t \right)$, where $\oy_2$ is the quadrature (defined with the 
same convention of Eq. (\ref{quad_def})) of the detection signal $\oy \tpar =\delta \ox \tpar + \ox_{\mathrm{imp}} \tpar$, 
composed by the signal of the mechanics $\delta\ox$ and
affected by the imprecision noise $ \ox_{\mathrm{imp}}$.

Substituting Eq.(\ref{quad_def}) in Eq.\eqref{eq_motion_1} and defining
the two quadratures of the force components as in Eq.(\ref{quad_def}) , 
the equations of motion for the two mechanical quadratures read:
\begin{align}
\delta \dot{\hat{X}}_1+\frac{\Gm}{2}\left(1+ \gs \right) \delta \Xo &=  \frac{\hat{F}_{\mathrm{th2}} +\hat{F}_{\mathrm{qba2}}}{2 m \Omega_\mathrm{m}},\\
\delta \dot{\hat{X}}_2+\frac{\Gm}{2}\left(1- \gs + \gfb \right) \delta \Xt &= - \frac{\hat{F}_{\mathrm{th1}} +\hat{F}_{\mathrm{qba1}}+m \Omega_\mathrm{m} \Gfb \ox_\mathrm{imp2}}{2 m \Omega_\mathrm{m}}, 
\end{align}
where  $\gfb=\Gfb/\Gm$.

These equations are conveniently solved in the Fourier domain, where $\omega$ is the frequency  in a frame rotating at $\Omega$
\begin{align}
\delta \Xot \om &= \chi_1 \om \left[ \tilde{{F}}_\mathrm{th2} \om + \tilde{{F}}_\mathrm{qba2} \om  \right],\\
\delta \Xtt \om &= - \chi_2 \om \left[ \tilde{{F}}_\mathrm{th1} \om + \tilde{{F}}_\mathrm{qba1} \om + m \Omega_\mathrm{m} \Gfb \oxt_\mathrm{imp2} \om  \right],
\end{align}
where 
\begin{align}
\chi_1 \om &= \frac{1}{2 m \Omega_\mathrm{m} \left[- i \omega + \frac{\Gm}{2} \left(1+\gs \right) \right]},
\\
\chi_2 \om &= \frac{1}{2 m \Omega_\mathrm{m} \left[- i \omega + \frac{\Gm}{2} \left(1-\gs+\gfb \right) \right]}.
\end{align}
The forces that drive the mechanical motion are delta-correlated, so that the $j$-th force contribution (with $j\in\{\mathrm{th},\mathrm{qba},\mathrm{ba}\}$) to the PSD of a motional quadrature $i\in\{1,2\}$ can be approximated to
\begin{equation}
{S}_{X_{i,j}} \om= \left| \chi_i \om \right|^2 \left[
{S}_{F_{j}} \left(\Omega_\mathrm{m}\right) + {S}_{F_{j}}   \left( - \Omega_\mathrm{m} \right) \right], \label{psd_general_formula}
\end{equation}
where the force PSD (${S}_{F_{j}} $) is considered flat in the frequency range
around the mechanical resonance.

\subsection{Noise contributions}

\subsubsection{Thermal contribution}
With ${S}_{F_\mathrm{th}} \left(\Omega_\mathrm{m}\right) + {S}_{F_\mathrm{th}}  \left( - \Omega_\mathrm{m} \right)= 2 m \Gm \hbar \Omega_\mathrm{m} \left(2 \bar{n} + 1\right) $,
the thermal contribution is easily evaluated to
\begin{gather}
{S}_{X_{i,\mathrm{th}}} \om= \frac{ \xzp ^2\Gm   \left(2 \bar{n} + 1\right)}{ \left[ \omega^2 + \left(\frac{\Gamma_{i}}{2} \right)^2 \right]},
\end{gather}
where  $\Go= \Gm \left(1+\gs \right)  $ and $\Gtw =\Gm \left(1-\gs+\gfb \right) $, and $\bar{n}$ is the average occupancy of the thermal bath at the mechanical frequency.

\subsubsection{Measurement back action contribution}
The fluctuating optical force due to the vacuum fluctuations
that enter the cavity reads:
\begin{equation}
\Fqba \tpar = \frac{\hbar g}{\xzp}\left[\delta \oa \tpar + \delta \oad \tpar \right],
\end{equation}
where $\delta \oa$ is the fluctuating component
of the cavity field $\oa$.
As such, the radiation pressure force exerted by the vacuum fluctuations on the oscillator can be written in the Fourier domain:
\begin{equation}
\Fqbat \om = \frac{\hbar \sqrt{\Gqba}}{\xzp} \left[\delta \oat_\mathrm{in} \om +\delta  \oatd_\mathrm{in} \om  \right],
\end{equation}
with $\Gqba= 4 g^2/ \kappa$.
The correlators of the vacuum noise read $\langle \delta\oat_\mathrm{in} \om \delta\oat_\mathrm{in} \omp \rangle = 
\langle \delta\oatd_\mathrm{in} \om \delta\oatd_\mathrm{in} \omp \rangle =
\langle \delta\oatd_\mathrm{in} \om \delta\oat_\mathrm{in} \omp \rangle = 0$, and
 $\langle \delta\oat_\mathrm{in} \om \delta\oatd_\mathrm{in} \omp  \rangle = \delta\left(
 \omega + \omega'\right)$.
Therefore, the contribution of the optical back action to the  PSD of the quadratures 
is
\begin{equation}
{S}_{X_{i,\mathrm{qba}}}  \om=  \frac{2 \xzp^2 \Gqba}{ \left[ \omega^2 + \left(\frac{\Gamma_{i}}{2} \right)^2 \right]}.
\end{equation}

\subsubsection{Feedback back action contribution}

To monitor the machanical motion we consider a homodyne detection scheme, optimized to record the
phase fluctuations of the signal field $\mathrm{\hat{Y}}\tpar = \frac{\delta \oa_\mathrm{out} \tpar - \delta \oad_\mathrm{out} \tpar}{2 i}$ (with $\oa_\mathrm{out}$ representing the output field of the cavity), which in the
bad cavity regime and in the Fourier domain read
\begin{equation}
\Yt \om = \left(\Yt_{\mathrm{in,loss}} \om\sqrt{1- \deteff}- \Yt_\mathrm{in} \om \sqrt{\deteff}\right) - \frac{\sqrt{\Gmeas} \delta\oxt(\omega)}{\xzp},
\end{equation}
where $\deteff$ is the detection efficiency, $\Gmeas=\deteff \Gqba$ is the measurement rate, $\Yt_{\mathrm{in,loss}}=\frac{\delta \oat_{\mathrm{in,loss}} \om  - \delta \oatd_{\mathrm{in,loss}} \om}{2 i}$ represents the noise that enters
via the loss channel of the detection and 
$\Yt_{\mathrm{in}}=\frac{\delta \oat_{\mathrm{in}} \om  - \delta \oatd_{\mathrm{in}} \om}{2 i}$ the noise that enters the cavity.

As such, the position measurement in the Fourier domain  reads
\begin{equation}
\oyt\om= \left( \Yt_\mathrm{in} \om \sqrt{\deteff}-\Yt_{\mathrm{in,loss}} \om\sqrt{1- \deteff}\right)\frac{\xzp}{\sqrt{\Gmeas}} +\delta \oxt \om. 
\end{equation}
The first term on the RHS of the equation is the
noise in the position detection $\oxt_\mathrm{imp}$ (imprecision noise).  
The heating term associated to the feedback is
\begin{equation}
\delta \Ffbt \om = m \Omega_\mathrm{m} \Gfb \oxt_\mathrm{imp2} \om, 
\end{equation}
where $\oxt_\mathrm{imp2} $ is the quadrature of the noise term as defined in Eq. (\ref{quad_def}).

From equation (\ref{psd_general_formula}), and the correlators of the input (quantum) noise, we find that the contribution 
to the quadrature PSD is
\begin{equation}
{S}_{X_\mathrm{2,fb}}  \om= \frac{\xzp^2 \Gfb^2}{8 \Gms} \frac{1}{ \left[ \omega^2 + \left(\frac{\Gamma_\mathrm{2}}{2} \right)^2 \right]}.
\end{equation}
No added noise comes from the correlation
between the measurement and the feedback backaction.

\subsubsection{Overall quadrature noise}
Summing up the individual contributions, we obtain for the two mechanical quadratures
\begin{align}
  {S}_{X_1}\om &    =\sum_j {S}_{X_{1,j}} 
                    = \frac{\xzp^2}{  \omega^2 + \left(\frac{\Gamma_\mathrm{1}}{2} \right)^2 }
                    \left( \Gm(2 \bar n+1) + 2\Gqba \right),\\
  {S}_{X_2}\om &    =\sum_j {S}_{X_{2,j}} 
                    = \frac{\xzp^2}{  \omega^2 + \left(\frac{\Gamma_\mathrm{2}}{2} \right)^2 }
                    \left( \Gm(2 \bar n+1) + 2\Gqba + \Gfb^2/8\Gmeas\right),
\end{align}
from which the variances given in the main manuscript are straightforwardly calculated.

\subsection{Homodyne PSDs}
The detected spectra of the two mechanical quadratures are computed evaluating
the PSDs of the  quadratures
of the detected homodyne signal as defined in Eq.~(\ref{quad_def}):
\begin{equation}
\Yt_{i} \om = \left(\Yt_{\mathrm{in,loss}} \om\sqrt{1- \deteff}- \Yt_\mathrm{in} \om \sqrt{\deteff}\right)_{i}- \frac{\sqrt{\Gmeas} \delta \tilde{X}_{i} \om }{\xzp},
\end{equation}
The PSDs for the two quadratures read
\begin{align}
  {{S}_{Y_1}\om} &=\frac{1}{2} + \frac{\Gmeas}{\xzp^2} \mathrm{S}_{X_1} \om,\\ 
  {{S}_{Y_2}\om} &=\frac{1}{2} + \frac{\Gmeas}{\xzp^2}\mathrm{S}_{X_2} \om - 
  \frac{\Gfb}{4} \frac{\Gamma_2}{\omega^2+\left(\frac{\Gamma_2}{2}\right)^2 },
  \label{eq:SY2}
\end{align}
where the first term is shot noise, the second term the transduced mechanical motion (including feedback-induced motion for the $Y_2$-quadrature), and the third term in equation (\ref{eq:SY2}) represents the feedback-induced correlation between motion and measurement noise (squashing).

\bibliographystyle{apsrev4-2}
\bibliography{references} 
